# Stream Control Transmission Protocol (SCTP): Robust and Efficient for Data Centre Applications

[1]Fatma Almajadub,     [2]Abdul Razaque,     [3]Eman Abdel Fattah
Computer Science Department
University of Bridgeport, CT, USA

**Abstract** — Due to rapid advancement in modern technology, one of the major concerns is the stability of business. The organizations depend on their systems to provide robust and faster processing of information for their operations. Efficient data centers are key sources to handle these operations. If the organization's system is not fully functional, the performance of organization may be impaired or clogged completely. With the developments of real-time applications into data centers for data communications, there is a need to use an alternative of the standard TCP protocol to provide reliable data transfer. Stream Control Transmission Protocol (SCTP) consists of several well built-in characteristics that make it capable to work efficiently with real-time applications. In this paper, we evaluate an optimized version of STCP. The optimized version of SCTP is tested against a non optimized version of STCP and TCP in a data center environment. Simulations of the protocols are carried using NS2 simulator.

*Index Terms*— SCTP, Data centers, TCP, simulation, performance.

## I. INTRODUCTION

The data centers represent the foundation of the Internet and computer services specially E-business service and high performance computing. Nowadays, the development of web services is based on the increased size and the complexity of the processed data. It is clear that the data centers continue to grow for higher performance and better availability. Hence, this remarkable growth in the data centers has motivated researchers to improve speed and capacity of data transfer [1].

Due to the heavy load of network traffic; TCP does not provide satisfactory performance in controlling congestion in data centers over the network. Thus, the network suffers from loss of confidential data. Although, TCP is deployed into the data centers, it does not have the capacity to control the huge amount of data [2]. There are several fuzzy questions about transparency of TCP protocol that is connected with IP protocol for supporting the applications of data centers. In data centers, there is a demand of high data rates, low latencies, high robustness and high availability

In spite of, the weaknesses of the standard TCP protocol to fit into the data centers are well known, it is impossible to do considerable changes on the standard TCP protocol [3, 4]. However, there are several alternative variants of TCP which are used in the areas where TCP cannot work. For example, SCTP which is a connection-oriented transport protocol that provides reliable stream oriented services similar to TCP. SCTP is especially designed to be used in situations where reliability and near real-time considerations are important as well as it is designed to run over existing IP/Ethernet infrastructure [5].

Moreover, SCTP was designed to support Signaling System number 7 (SS7) layers which have unique features that are suitable for data centers. Also, SCTP has many promising features including flexibility, robustness, and extensibility. Therefore, SCTP protocol is a better choice for data centers [6, 7].

In this paper, we examine using SCTP into data centers. Furthermore, we evaluate the effect of optimizing some parameters on the overall performance. The paper is organized as follows: in section II, we present the features of SCTP for data center requirements. In section III, Implementation of LK-SCTP is introduced. In section IV, simulation results are presented. Finally, section V provides conclusions and future work.

## II. FEATURES OF SCTP FOR DATA CENTER REQUIREMENTS

Although the standard TCP protocol has many features, it was not designed to be used for the data centers. Consequently, the need for a better protocol such as SCTP is established. SCTP adopts congestion window/flow control scheme of TCP except for some minor differences [8, 9]. Therefore, SCTP is identical to the standard TCP protocol with regards to congestion and flow control.

On the other hand, SCTP has provided many improvements over the standard TCP as follows:

a) Multi-streaming: SCTP connection can have multiple streams; each of them specifies a logical channel as shown in Figure 1. Although the flow and congestion control are still on the basis of each connection, the streams can be exploited for many purposes such as assigning higher priority to messages [10, 11].

---

[1] Fatma Almajadub: Master student at University of Bridgeport, Computer Science and Engineering Department, University of Bridgeport, CT-06604, USA, (falmajad@bridgeport.edu).
[2] Abdul Razaque: Research Scientist at University of Bridgeport, CT-06604, USA, (arazaque@bridgeport.edu).
[3] Eman Abdel Fattah: Adjunct Instructor at University of Bridgeport, Computer Science and Engineering Department, University of Bridgeport, CT-06604, USA, (eman@bridgeport.edu).

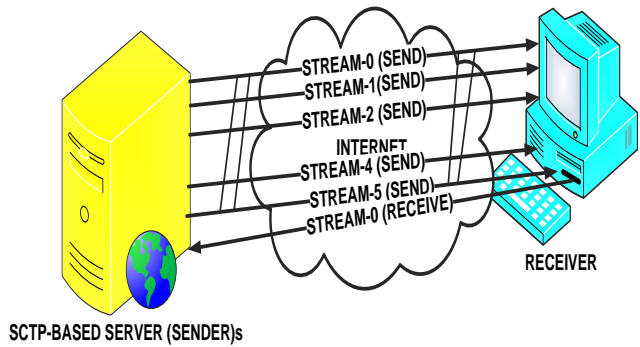

Figure 1: Multi-streaming process of SCTP

b) Multi-homing: SCTP connection can define multiple "endpoints" on each end of the connection that increases the capability to handle errors. If primary connection fails, then the sender selects alternate connection for forwarding data until it is restored as shown in Figure 2.

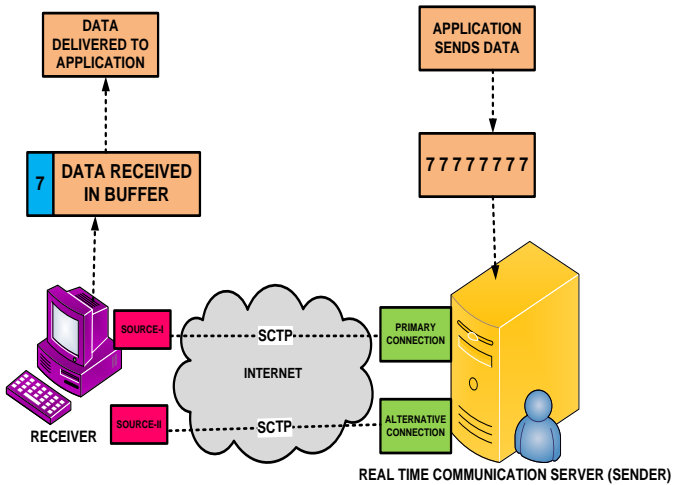

Figure 2: Multi-homing process of SCTP

c) One of the promising features of SCTP is the capability to handle denial of service attacks. SCTP connection includes a 4-way handshaking process to prevent propagation of any message at the endpoint until it has ensured that the other end is interested in setting up a connection as shown in Figure 3 [12].

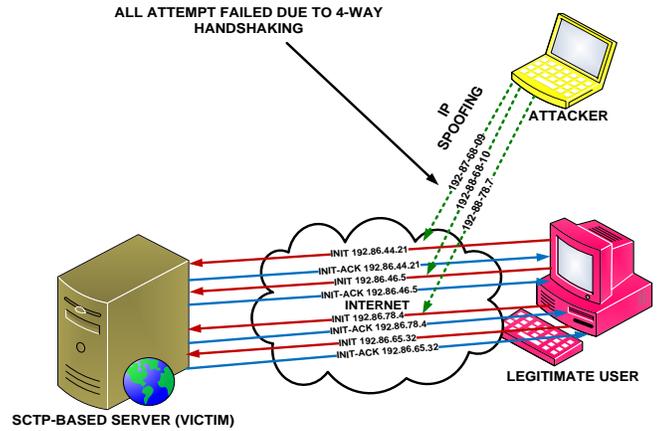

Figure 3: 4-way handshaking process of SCTP

d) In-order delivery: SCTP stream provides well organized in-order delivery which reduces latency [13].
e) Robust connection: SCTP connection maintains a verification tag that is provided for each subsequent data transfer so that it is robust against tapping and errors. This feature is vital within the data center for transferring high data rates [14, 15].

## III. IMPLEMENTATION OF LK-SCTP

In this section we discuss the LK-SCTP Implementation. We choose Linux Kernel (LK-SCTP) implementation as a basis for evaluating performance enhancements in section IV. Figure 4 shows the approach used in LK-SCTP to chunk the messages at the senders' side and chunk bundling at the receiver side. The approach works as follows:

I. The message contains the list of chunks. LK-SCTP maintains three data structures to manage the chunks.
II. When all the chunks are acknowledged by the remote endpoint, the first data structure is free.
III. LK-SCTP uses the two other data structures to manage each chunk as follows:
   a. The first structure contains the actual chunk buffers and the chunk header.
   b. The second structure contains pointers to the chunk buffers and other data.
IV. Other small data structures are maintained by the implementation. The chunk is copied to the final buffer after it is processed by many procedures and routines. Before LK-SCTP copies variables and their values to the final destination, it initializes the local variables with values.

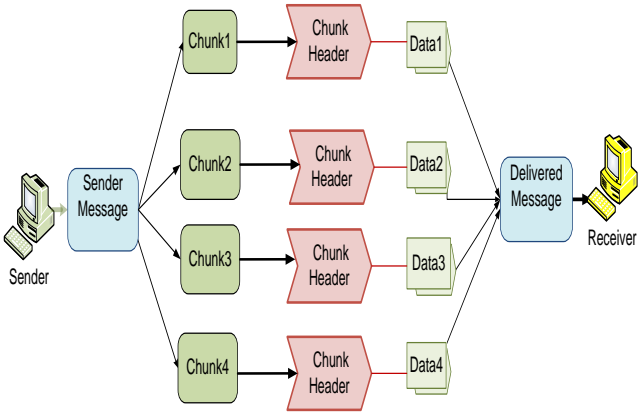

Figure 4: Process of sending message

Three Memory to Memory (M2M) copies are generated before data is transmitted. The first copy is retrieving the data from the user buffer as well as writing in the data message structure. The second is used to pass control to NIC by bundling chunks into MTU packet. The third copy is a direct memory access (DMA) of data into the NIC buffers.

## IV. SIMULATION RESULTS

A comparison between the performance of TCP, SCTP without optimization, and SCTP with optimization is presented in this section. The CPU Utilization, Goodput, and Throughput are evaluated. Table 1 shows the parameters that we have tuned for the NS-2 Simulator in the case of optimized SCTP.

### A. Performance Impact of Optimizations of SCTP

Figure 5 compares the average CPU utilization of TCP, SCTP without optimization, and SCTP with optimization for 12 KB packets. At a steady state after 20 minutes, SCTP without optimization performs worse than TCP. Furthermore, SCTP with optimization performs better than both TCP and SCTP without optimization.

CPU utilization can be obtained as follows:

CPU Utilization= ($USCPU_1$ + $USCPU_2$ + $USCPU_3$ …, $USCPU_n$) / Number of USCPU

Where,

USCPU is the utilized sample for CUP that is based on Idle CPU, which is given as below:

USCPU= 100 %-(% time consumed in idle task)- ( CPU %Instructions per fetch)----------------(1)

Assume, we have 6% time consumed for idle task and 25% CUP for instructions per fetch in $USCPU_1$. Similarly, 4% time consumed for idle task and 20% CUP for instructions per fetch in $USCPU_2$ and we have 7% time consumed for idle task and 30% CUP for instructions per fetch in $USCPU_3$. Thus, USCPU1 = 100-6-25 = 69
USCPU2 =100-4-20=76
USCPU3=100-7-30= 63
Therefore,

Table 1: Parameters used in NS-2 Simulator for Optimized SCTP

| | |
|---|---|
| Debug Mask | 1 |
| Debug File Index | 0 |
| MTU | 1500 |
| Data Chunk Size | 512 and 1468 |
| Number of out streams | 1 |
| CMT congestion window | 1 |
| CMT Del Acknowledgement | 1 |
| RTX congestion window | 4 |
| Heart Beat Timer | 0 |
| Initial Receiving window | 65536 |
| Queue size limit | 50 |
| HB .interval | 25 seconds |
| Maximum Initial Retransmits | 9 attempts |
| RTO.Initial | 4 seconds |
| RTO.Max | 60 seconds |
| RTO.Min | 1 second |
| RTO.Beta | 1/4 |
| RTO.Alpha | 1/8 |
| Association Maximum Retransmission | 10 attempts |
| Valid cookie life | 50 seconds |
| Path Maximum Retransmission | 6 attempts (per destination address) |
| Application Buffer size | 0 |
| Send Buffer Size | 0 |
| Channel type | Wireless Channel |
| Drop Tail | 5 Mb 200ms |
| Simulation time | 400 seconds |
| Packet size | 1024 bytes |
| Application | ftp |
| Burst time | 0.5 second |
| No: of changes | 10 |
| Radio-propagation model | Two Ray Ground |
| Network interface type | OFDM |
| MAC type | Mac/802_16/BS |
| Link Layer type | Logical Link |
| Interface queue type | Drop Tail/Priority Queue |
| Pause time | 3 seconds |

CPU Utilization= (USCPU$_1$ + USCPU$_2$ + USCPU$_3$ …, USCPU$_n$) / Number of USCPU----------------------------(2)
Substitute the values:
CPU Utilization= (69+76+63) /3
CPU Utilization= 69.33

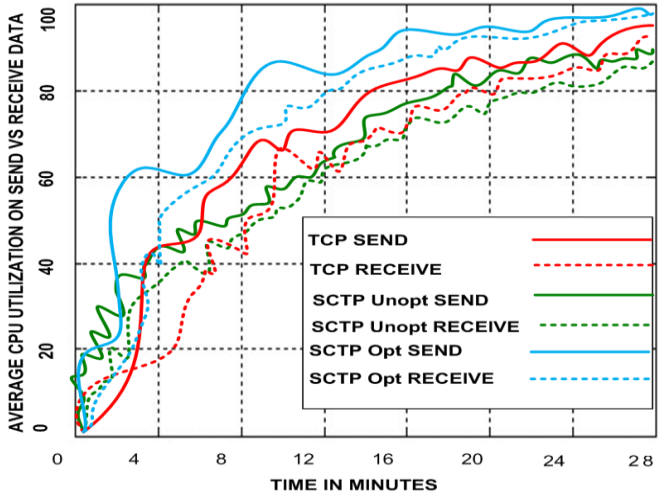
Figure 5: Average CPU utilization for 12 KB packets

Figure 6 shows the data rate versus the number of connections. It shows that the three protocols scale up with the number of connections. At a steady state after 5 connections, SCTP without optimization performs as comparable as TCP. Furthermore, SCTP with optimization performs better than both TCP and SCTP without optimization.

Average throughput can be achieved as follows:
$B_c = B_{E-1} * RTT + P_k /RTT+T_k$
Where,
Bandwidth= B, Band width at acknowledged segments= $B_{E-1}$, Round trip time= RTT, Packet Size= $P_k$ and Current Bandwidth= $B_c$
Maximum Congestion window size (W) = RTT *B/ $P_k$ --(2)
Thus,
Throughput= W* MSS/RTT-------------------------------(3)
Here, MSS is the maximum segment size.

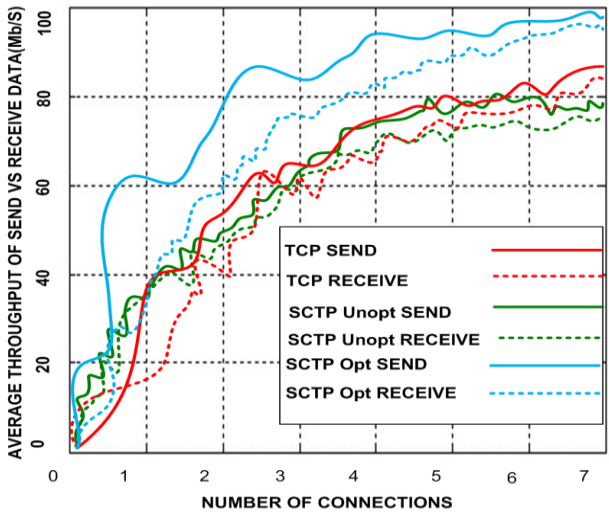
Figure 6: Throughput scaling with multiple connections

Figure 7 shows the SCTP goodput for small packets which are around 128KB. Goodput is defined as the average amount data received by the receiver per unit time that are not retransmissions [16]. As shown in Figure 7, the goodput is almost stable over time. Furthermore, the goodput of SCTP with optimization and SCTP without optimization is better than the goodput of TCP.
Goodput can be calculated as follows:
GP = ACK$_{seg}$ * 100/ SENT$_{seg}$ --------------------------------(4)
Here,
ACK$_{seg}$ = Acknowledged segments and $_{SENTseg}$ = Sent segment

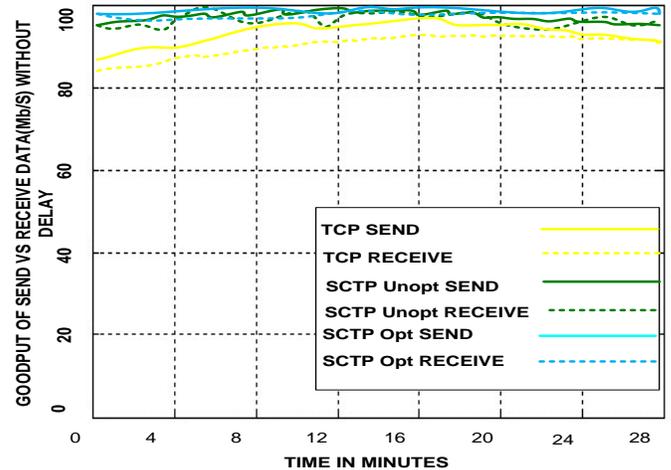
Figure 7: Goodput comparisons with 128 Bytes packets

Figure 8 shows the packet loss rate versus the data rate. The SCTP protocol achieves better results than TCP. For a particular data rate, the percentile of packet loss is less for SCTP than TCP.

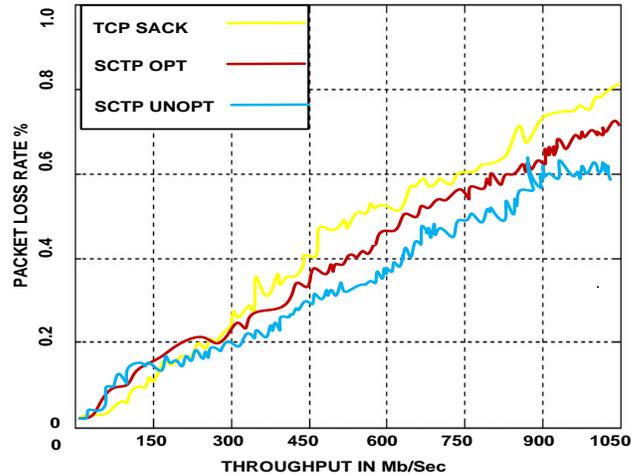
Figure 8: Throughput comparisons versus the packet loss

V. CONCLUSION AND FUTURE WORK

In this paper, we have studied the features of SCTP from data center point of view. CPU Utilization, Throughput, and goodput of SCTP and TCP are examined. SCTP implementations with optimization and without optimizations are considered in our simulation. It is found that SCTP without optimization performs worse than TCP in most cases. An optimized SCTP implementation performs better than TCP and SCTP without optimization. In the future we plan to study the effect of stream priories and topology of devices on the performance of STCP in a data center environment.